\documentclass[a4paper,oneside,12pt]{article}

\newtheorem{dfn}{Definicja}[section]
\newtheorem{tw}[dfn]{Theorem} 
\newtheorem{prop}[dfn]{Proposition}

\newtheorem{rem}[dfn]{Remark}
\usepackage{verbatim}
\usepackage{array}
\usepackage{enumerate}
\usepackage{amssymb} 
\usepackage{amsmath}
\usepackage{fancyhdr}
\usepackage{euscript}
\usepackage{latexsym}

\author{Micha\l \ Barski \\
\small Faculty of Mathematics and Computer Science, University of
Leipzig\\
\small Faculty of Mathematics, Cardinal Stefan Wyszy\'nski University in Warsaw\\
\small{\small{\it Michal.Barski@math.uni-leipzig.de}}}

\title{\bf Quantile hedging for basket derivatives
}

\begin{document}
\baselineskip=16pt
\maketitle
\date
\begin{abstract}
The problem of quantile hedging for basket derivatives in
the Black-Scholes model with correlation is considered. Explicit
formulas for the probability maximizing function and the cost
reduction function are derived. Applicability of the results for the
widely traded derivatives as digital, quantos, outperformance and
spread options is shown.
\end{abstract}

\noindent
\begin{quote}
\noindent \textbf{Key words}: quantile hedging, basket derivatives, correlated assets.

\textbf{AMS Subject Classification}: 91B30, 91B24, 91B70,

\textbf{JEL Classification Numbers}: G13,G10.
\end{quote}

\section{Introduction}
As recent events on the market have shown the risk appearing in
pricing of financial contracts should be more thoroughly surveyed.
Although the problem of minimizing risk is widely studied in the
literature, the great majority of the results do not meet the
expectations of practitioners who are interested in straightforward
applications. This paper is concerned with the issue of risk
analysis for the basket derivatives and provides explicit
computing methods for the risk parameters.

The risk is measured by the possibility of a partial hedging of the
payoff. Thus our approach is based on the idea of quantile hedging 
which was introduced in \cite{FL1} and later developed in various
directions, see for instance \cite{CvitanicSpivak},
\cite{MelnikovKrutchenko}, \cite{MB1}, \cite{MB2}. Let us briefly
sketch a general concept. Denote by $H$ a contingent claim and
assume that the arbitrage free pricing method indicates its price
$p(H)$. This means that if the investor has an initial endowment
$x\geq p(H)$ then he is able to follow some trading strategy such
that his portfolio hedges $H$ with probability $1$. If this is the
case, then $x$ carries no risk and {\it the probability maximizing
function} $\Phi_1$ equals $1$, i.e. $\Phi_1(x)=1$. On the other
hand, if $x<p(H)$ then the shortfall probability is strictly greater than zero for each trading strategy and then
$\Phi_1(x)<1$. The grater the probability of shortfall is the smaller
the value $\Phi_1(x)$ is. Thus the function $\Phi_1$ can be viewed
as a measure of the risk sensitivity to the price reduction of the option.
There is also another aspect of the problem. Assume
that the hedger is willing to accept some risk measured by the
shortfall probability in order to reduce initial cost. 
He chooses a number $\alpha\in[0,1]$ and searches for
a minimal initial capital $\Phi_2(\alpha)$ which allows to find a
strategy such that the probability of the shortfall is
smaller then $1-\alpha$. Thus if the hedger accepts no risk, i.e.
$\alpha=0$, then the minimal cost required to replicate $H$ is just
$p(H)$. In this case {\it the cost reduction function} satisfies
$\Phi_2(0)=p(H)$. However, if $\alpha>0$ then $\Phi_2(\alpha)<p(H)$
and the function $\Phi_2$ enables us to view the effect how the risk
acceptance affects the cost reduction of the option. Recall the
numerical example from \cite{FL1} p. 261 which shows that
$\Phi_2(0,05)=0,59\cdot p(H)$ for a call option with certain
parameters. This means that the acceptance of a $5\%$ margin of risk
reduces the hedging cost by $41\%$. This shows that quantile
hedging is an attractive tool for the risk analysis
and should be taken into account by traders.

The basic problem, however, is to determine functions $\Phi_1$ and
$\Phi_2$ for specific derivatives. There are only a few examples in
the literature where they are explicitly found. In \cite{FL1}
explicit formulas are given for the most important case of a call
option in a classical Black-Scholes model. The method can be
mimicked to obtain formulas for the put option. The idea is based on
reducing the original dynamic problem to the static one which can be
solved with methods used in the theory of statistical tests.
Since the market was complete the solution of the static problem
could be obtained, via Neyman-Pearson lemma, by indicating a
non-randomized test for the appropriate probability measures.
The Neyman-Pearson lemma can be generalized for the case of composite hypotheses, i.e.
when measures are replaced by the families of measures, see
\cite{Cvitanic} where the solution in the abstract
form is presented. However, straightforward applicability of this result towards incomplete markets
seems to be questionable. This paper is devoted to determining
functions $\Phi_1$ and $\Phi_2$ for the basket derivatives
in the Black-Scholes framework with correlation. As the market is complete, we follow the same general
method as in \cite{FL1}, but we find the solutions explicitelly using specific features of the model.
More preciselly, we show that the
original problem can be reduced to that of finding another two
deterministic functions $\Psi_1, \Psi_2$ depending on $H$, which turned out to be regular, i.e. continuous and strictly 
monotone if $H$ is of a reasonable form, see Propsosition \ref{proposition o Psi_1} and Proposition \ref{proposition o Psi_2}.
Then, roughly speaking,
$\Phi_1=\Psi_1\circ\Psi^{-1}_2$ and $\Phi_2=\Psi_2\circ\Psi^{-1}_1$;
for a precise formulation see Theorem \ref{TH MAIN}. In the one dimensional case 
when $H$ is a call option the result covers the above mentioned example from \cite{FL1}.
We also determine explicit forms of $\Psi_1$ and $\Psi_2$ for commonly
traded derivatives, see Section \ref{Quantile hedging in two
dimensional model} and its subsections. As $\Psi_1, \Psi_2$ are rather of a complicated form, the inverse
functions can not be given by analytic formulas but can be determined with the use of numerical methods. 
Thus a great advantage of our results is that they can be used in practice.

The paper is organized as follows. In Section \ref{The model} we
briefly recall the multidimensional Black-Scholes model and
formulate the problem strictly. Section \ref{Description of the
method} contains the main result - Theorem \ref{TH MAIN} which is
proceeded by a general discussion on the results from \cite{FL1} and
the Neyman-Pearson technique. The method established in Theorem
\ref{TH MAIN} is used in Section \ref{Quantile hedging in two
dimensional model} for calculating the functions $\Psi_1$, $\Psi_2$
for two assets derivatives which are widely traded, that is for
digital option, quanto domestic, quanto foreign,
outperformance and spread options.

\newpage

\section{The model}\label{The model}
Let $(\Omega,\mathcal{F}_t,t\in[0,T], P)$ be a fixed probability
space with filtration. The prices of $d$ shares are given by the
Black-Scholes equations
\begin{gather*}
dS^i_t=S^i_t(\alpha_i dt+\sigma_{i}dW^i_t), \qquad i=1,2,...,d, \
t\in[0,T],
\end{gather*}
where $\alpha_i\in\mathbb{R}$, $\sigma_i>0, \ i=1,2,...,d$ and
$W_t=(W^1_t,W^2_t,...,W^d_t), t\in[0,T]$, is a sequence of standard
Wiener processes adapted to $\{\mathcal{F}_t; t\in[0,T]\}$ with the
correlation matrix $Q$ of the form
\begin{gather*}
Q=\left[\begin{array}{rrrrr}
1&\rho_{1,2}&\rho_{1,3}&\ldots&\rho_{1,d}\\
\rho_{2,1}&1&\rho_{2,3}&\ldots&\rho_{2,d}\\
\vdots&\vdots&\vdots&\vdots&\vdots\\
\rho_{d,1}&\rho_{d,2}&\rho_{d,3}&\ldots&1\\
\end{array}\right],
\end{gather*}
where
\begin{gather*}
\rho_{i,j}=cor\left\{W^i_1,W^j_1\right\}, \quad i,j=1,2,...,d.
\end{gather*}
We assume that $Q$ is positive definite. The process given above will
be called a $Q$-Wiener process. The trader can invest his money in
stocks as well as put it on a savings account which dynamics is
given by
\begin{gather*}
dB_t=rB_tdt, \quad t\in[0,T],
\end{gather*}
with $r$ standing for a constant short rate.

\begin{rem} The most common approach for the description of the market
is based on a sequence of independent Wiener processes, see for
instance a classical textbook \cite{Karatzas}. It can be shown that
the model described above is equivalent to the model with $d$
independent Wiener processes and the $d\times d$ diffusion matrix
with constant coefficients. We work with a correlated
Wiener process because it is more convenient for later calculations.
Let us also mention that parameters in such model can be easily
estimated from data, see \cite{Glasserman} p.104.
\end{rem}

Let us now briefly characterize a martingale measure of the model,
i.e. a measure $\tilde{P}$ which is equivalent to $P$ such that the
discounted price processes $\hat{S}_t^i:=e^{-rt}S^i_t, i=1,2,...,d$
are martingales. The following is a version of Theorem 10.14
in \cite{DaPrato-Zabczyk} adapted to our finite dimensional setting.
\begin{tw}\label{tw Girsanova}
Let $\varphi$ be a predictable process taking values in
$\mathbb{R}^d$ satisfying
\begin{gather*}
\mathbf{E}\left(
e^{\int_{0}^{T}(Q^{-\frac{1}{2}}\varphi_t,dW_t)-\frac{1}{2}\int_{0}^{T}\mid\varphi_t\mid^2dt}\right)=1.
\end{gather*}
Then the process
\begin{gather*}
\widetilde{W}_t=W_t-\int_{0}^{t}Q^{\frac{1}{2}}\varphi_sds, \qquad
t\in[0,T],
\end{gather*}
is a $Q$- Wiener process with respect to the measure $\widetilde{P}$
with a density
\begin{gather*}
\frac{d\widetilde{P}}{dP}=e^{\int_{0}^{T}(Q^{-\frac{1}{2}}\varphi_t,dW_t)-\frac{1}{2}\int_{0}^{T}\mid\varphi_t\mid^2dt}.
\end{gather*}
\end{tw}

It can be shown that each measure equivalent to $P$ can be
characterized by a density process
\begin{gather}\label{gestosc miary rownowaznej}
Z_t:=e^{\int_{0}^{t}(Q^{-\frac{1}{2}}\varphi_s,dW_s)-\frac{1}{2}\int_{0}^{t}\mid\varphi_s\mid^2ds},
\qquad t\in[0,T],
\end{gather}
for some predictable $\mathbb{R}^d$- valued process $\varphi$. The
process $\hat{S}^{i}$ is a  $\widetilde{P}$  martingale if and
only if $\hat{S}^iZ$ is a $P$ martingale. Thus the measure
$\widetilde{P}$ can be determined by finding a process $\varphi$ in
\eqref{gestosc miary rownowaznej} such that $\hat{S}^iZ,
i=1,2,...,d$ are $P$ martingales. Simple calculations based on
the It\^o formula yield
\begin{gather*}
\varphi_t=-Q^{-\frac{1}{2}}\left[\frac{\alpha-r\mathbf{1}_d}{\sigma}\right]:=-Q^{-\frac{1}{2}}\left[\begin{array}{r}
\frac{\alpha_1-r}{\sigma_1}\\
\frac{\alpha_2-r}{\sigma_2}\\
\vdots\\
\frac{\alpha_d-r}{\sigma_d}\\
\end{array}\right], \qquad t\in[0,T].
\end{gather*}
The martingale measure $\widetilde{P}$ is thus unique and given by
the density process
\begin{gather}\label{gestosc miary martyngalowej - ogolnie}
\tilde{Z}_t:=e^{-(Q^{-1}[\frac{\alpha-r\mathbf{1}_d}{\sigma}],W_t)-\frac{1}{2}\mid
Q^{-\frac{1}{2}}[\frac{\alpha-r\mathbf{1}_d}{\sigma}]\mid^2t},
\qquad t\in[0,T].
\end{gather}
Moreover, it follows from Theorem \ref{tw Girsanova} that the
process
\begin{gather*}
\widetilde{W}_t:=W_t+\frac{\alpha-r\mathbf{1}_d}{\sigma} \ t, \qquad
t\in[0,T],
\end{gather*}
is a $Q$- Wiener process under $\widetilde{P}$. The dynamics of the
prices under the measure $\widetilde{P}$ can be written as
\begin{gather*}
dS^i_t=S^i_t(rdt+\sigma_id\widetilde{W}^i_t), \quad i=1,2,...,d.
\end{gather*}
The wealth process with the initial endowment $x$ and the trading
strategy $\pi$ is defined by
\begin{gather*}
X^{x,\pi}_t:=\pi^{0}_{t}B_t+\sum_{i=1}^{d}\pi^{i}_{t}S^{i}_{t},
\quad t\in[0,T].
\end{gather*}
and assumed to satisfy $X^{x,\pi}_0=x$. All strategies are assumed to be admissible, i.e.
$X^{x,\pi}_{t}\geq0$ for each $t\in[0,T]$ almost surely and
self-financing, i.e.
\begin{gather*}
dX^{x,\pi}_{t}=\pi^{0}_{t}dB_t+\sum_{i=1}^{d}\pi^{i}_{t}dS^{i}_{t},
\quad t\in[0,T].
\end{gather*}
A {\it contingent claim}, representing future random payoff,
is a random variable $H\geq0$ measurable wrt. $\mathcal{F}_T$.  A {\it hedging
strategy} against $H$ is a pair $(x,\pi)$ such that
\begin{gather*}
P(X^{x,\pi}_T\geq H)=1.
\end{gather*}
A {\it replicating strategy} is a pair $(x,\pi)$ such that
\begin{gather*}
P(X^{x,\pi}_T= H)=1.
\end{gather*}
A {\it price} of $H$ is defined by
\begin{gather*}
p(H):=\inf\left\{x: \exists \pi \ \text{s.t.} \ P(X^{x,\pi}_T\geq
H)=1\right\}
\end{gather*}
and, due to the fact that the market is complete, it follows from
the general theory that $p(H)=\mathbf{\tilde{E}}[e^{-rT}H]$, where
the expectation is calculated under the measure $\widetilde{P}$.

If $x<p(H)$ then $P(X_T^{x,\pi}\geq H)<1$ for all $\pi$ and the
question under consideration is to find a strategy maximizing the
probability of successful hedge, i.e.
\begin{gather}\label{problem 1 max prob}
P(X_T^{x,\pi}\geq H)\underset{\pi}{\longrightarrow}\max.
\end{gather}
We will refer the corresponding function
$\Phi_1:[0,+\infty)\longrightarrow[0,1]$ given by
\begin{gather*}
\Phi_1(x):=\max_{\pi}P(X_T^{x,\pi}\geq H),
\end{gather*}
as the {\it maximal probability function}. If there exists $\hat{\pi}$
such that $P(X^{x,\hat{\pi}}_T\geq H)=\Phi_1(x)$ then it will be called {\it
the probability maximizing strategy for $x$}.

We also consider the problem of cost reduction. Let
$\alpha\in[0,1]$ be a fixed number describing the level of shortfall
risk accepted by the trader. Then we are searching for a minimal
initial cost such that there exists a strategy with the probability
of successful hedge exceeding $1-\alpha$, i.e.
\begin{gather}\label{problem 2 min cost}
x\longrightarrow \min; \quad \text{$\exists$ $\pi$ \ s.t.} \quad
P(X_T^{x,\pi}\geq H)\geq 1-\alpha.
\end{gather}
The {\it cost reduction function}
$\Phi_2:[0,1]\longrightarrow[0,p(H)]$ is thus defined by
\begin{gather*}
\Phi_2(\alpha):=\min\left\{x: \exists \pi  \ \text{s.t.} \
P(X_T^{x,\pi}\geq H)\geq 1-\alpha\right\}.
\end{gather*}
If there exists $\hat{\pi}$ such that
$P(X^{\Phi_2(\alpha),\hat{\pi}}_T\geq H)\geq 1-\alpha$ then it will be
called {\it the cost minimizing strategy for $\alpha$}.

In the sequel we study the problem of determining the functions
$\Phi_1$ and $\Phi_2$ for the contingent claim $H$ of a general
form. Then in Section \ref{Quantile hedging in two dimensional
model} specific payoffs are examined.

\section{Main results}\label{Description of the method}
In this section we present a general method of determining functions
$\Phi_1$ and $\Phi_2$. Let us start with the auxiliary problems
which can be solved via the Neyman-Pearson lemma.

 Assume that we are given two probability measures $P_1$,
$P_2$ with strictly positive density $\frac{dP_1}{dP_2}$ and
consider two types of optimizing problems
\begin{gather}\label{pierwszy NP ogolny}
\begin{cases} \ P_1[A]\longrightarrow
\max,\\
P_2[A]\leq x,
\end{cases}
\end{gather}
\begin{gather}\label{drugi NP ogolny}
\begin{cases} \ P_1[B]\geq 1-\alpha\\
P_2[B]\longrightarrow \min,
\end{cases}
\end{gather}
where $\alpha,x\in[0,1]$ are fixed constants. Problem
$\eqref{pierwszy NP ogolny}$ is a classical one appearing in the
statistical hypotheses testing. Recall, that if there exists a
constant $c\geq0$ such that $P_2(\frac{dP_1}{dP_2}\geq c)=x$ then
the set
\begin{gather*}
\tilde{A}:=\left\{\frac{dP_1}{dP_2}\geq c\right\}
\end{gather*}
is a solution of $\eqref{pierwszy NP ogolny}$. It is not surprising
that the solution of the problem $\eqref{drugi NP ogolny}$ is of a
similar form. For the convenience of the reader we prove the
following.
\begin{prop}
If there exists a constant $c\geq0$ satisfying
$P_1(\frac{dP_2}{dP_1}\leq c)=1-\alpha$ then the set
\begin{gather*}
\tilde{B}:=\left\{\frac{dP_2}{dP_1}\leq c\right\}
\end{gather*}
is a solution of the problem $\eqref{drugi NP ogolny}$.
\end{prop}
{\bf Proof:} Let $B$ be an arbitrary set satisfying $P_1(B)\geq
1-\alpha$. We will show that $P_2(B)\geq P_2(\tilde{B})$. The
following estimation holds.
\begin{align*}
P_2(B)&-P_2(\tilde{B})=\int_{\Omega}(\mathbf{1}_{B}-\mathbf{1}_{\tilde{B}})dP_2
=\int_{\{\frac{dP_2}{dP_1}\leq c\}}(\mathbf{1}_{B}-\mathbf{1}_{\tilde{B}})dP_2\\[1ex]
&+\int_{\{\frac{dP_2}{dP_1}>c\}}(\mathbf{1}_{B}-\mathbf{1}_{\tilde{B}})dP_2\geq c \int_{\{\frac{dP_2}{dP_1}\leq c\}}(\mathbf{1}_{B}-\mathbf{1}_{\tilde{B}}) dP_1+c \int_{\{\frac{dP_2}{dP_1}>c\}}\mathbf{1}_{B} dP_1\\[1ex]
&=c\left(\int_{\Omega}\mathbf{1}_{B}dP_1-\int_{\Omega}\mathbf{1}_{\tilde{B}}dP_1\right)=c(P_1(B)-P_1(\tilde{B}))\\[1ex]
&\geq c \Big(P_1(B)-(1-\alpha)\Big)\geq0.
\end{align*}
\hfill $\square$

\vskip2ex

\noindent Let us notice that both optimal sets $\tilde{A}$,
$\tilde{B}$ have a similar form
\begin{gather}\label{postac optymalnych zbiorow}
\left\{\frac{dP_1}{dP_2}\geq c\right\},
\end{gather}
with suitable constants $c\geq0$. More precisely, for $\tilde{A}$ the
constant $c$ is s.t.
\begin{gather}\label{warunek ogolny na c(x)}
P_2\left(\frac{dP_1}{dP_2}\geq c\right)=x
\end{gather}
and for $\tilde{B}$ is s.t.
\begin{gather}\label{warunek ogolny na c(alfa)}
P_1\left(\frac{dP_1}{dP_2}\geq c\right)=1-\alpha.
\end{gather}
Now, come back to the initial problem of determining functions
$\Phi_1$, $\Phi_2$. Let us start with presenting two auxiliary
results which are nonrandomized versions of Theorems 2.34 and 2.42
in \cite{FL1}.

\begin{tw}\label{TH 1 max prob}
Let $x\geq0$. If $\tilde{A}$ is a set solving the problem
\begin{gather}\label{twierdzenie o tilde A}
\begin{cases} \ P[A]\longrightarrow
\max,\\
\mathbf{\tilde{E}}[e^{-rT}H\mathbf{1}_A]\leq x,
\end{cases}
\end{gather}
then $\Phi_1(x)=P(\tilde{A})$ and the probability maximizing
strategy for $x$ is that one replicating the payoff
$H\mathbf{1}_{\tilde{A}}$.
\end{tw}
Let us notice that if $x\geq p(H)$ then $\tilde{A}=\Omega$ and thus
$\Phi_1(x)=1$. Moreover, if \eqref{twierdzenie o tilde A} has a
solution for every $x\geq0$, then the function $\Phi_1$ is
increasing.
\begin{tw}\label{TH 2 min cost}
Let $\alpha\in[0,1]$ be a fixed number. If $\tilde{B}$ is a set
solving the problem
\begin{gather}\label{twierdzenie o tilde B}
\begin{cases} \ P[B]\geq 1-\alpha,\\
\mathbf{\tilde{E}}[e^{-rT}H\mathbf{1}_B]\longrightarrow\min,
\end{cases}
\end{gather}
then
$\Phi_2(\alpha)=\mathbf{\tilde{E}}[e^{-rT}H\mathbf{1}_{\tilde{B}}]$
and the cost minimizing strategy for $\alpha$ is that one
replicating the payoff $H\mathbf{1}_{\tilde{B}}$.
\end{tw}
Notice that $\Phi_2(0)=p(H)$ and if \eqref{twierdzenie o tilde B}
has a solution for each $\alpha\in[0,1]$ then $\Phi_2$ is
decreasing.

Now apply the method of solving the problems $\eqref{pierwszy NP
ogolny}$ and $\eqref{drugi NP ogolny}$ to \eqref{twierdzenie o tilde
A} and \eqref{twierdzenie o tilde B}. Notice that \eqref{twierdzenie
o tilde A} and \eqref{twierdzenie o tilde B} can be reformulated to
the following form
\begin{gather}\label{TH 1 max prob (2)}
\begin{cases} \ P[A]\longrightarrow
\max,\\
P^\ast(A)\leq \frac{x}{\mathbf{\tilde{E}}[e^{-rT}H]},
\end{cases}
\end{gather}
and
\begin{gather}\label{TH 2 min cost (2)}
\begin{cases} \ P[B]\geq 1-\alpha\\
P^{\ast}(B)\longrightarrow \min,
\end{cases}
\end{gather}
where $P^\ast$ is a probability measure given by the density
\begin{gather*}
\frac{dP^\ast}{d\widetilde{P}}=\frac{H}{\mathbf{\tilde{E}}[H]}.
\end{gather*}
In view of \eqref{postac optymalnych zbiorow} we are searching for
the solutions $\tilde{A}$, $\tilde{B}$ to \eqref{TH 1 max prob (2)},
\eqref{TH 2 min cost (2)} in the family of sets
\begin{gather*}
\left\{\frac{dP}{dP^\ast}\geq
c\right\}=\left\{\frac{dP}{d\widetilde{P}}\frac{d\widetilde{P}}{dP^\ast}\geq
c\right\}=\left\{\tilde{Z}^{-1}_{T}\geq c\frac{H}{\mathbf{\tilde{E}}[H]}\right\};
\quad c\geq0,
\end{gather*}
where $\tilde{Z}_T$ is given by \eqref{gestosc miary martyngalowej -
ogolnie}. Denoting, for the sake of simplicity, the constant
$\frac{c}{\mathbf{\tilde{E}}[H]}$ by $c$ we see that the
optimal sets $\tilde{A}$, $\tilde{B}$ are of the form
\begin{gather}\label{postac A_c}
A_c:=\left\{\tilde{Z}_T^{-1}\geq cH\right\},
\end{gather}
where, by \eqref{warunek ogolny na c(x)} and \eqref{warunek ogolny
na c(alfa)},  $c$ is s.t.
\begin{gather}\label{warunek na c(x) (2)}
P^{\ast}(A_c)=\frac{x}{\mathbf{\tilde{E}}[e^{-rT}H]} \quad
\text{for} \ \tilde{A},
\end{gather}
and
\begin{gather}\label{warunek na c(alfa) (2)}
P(A_c)=1-\alpha \quad \text{for}  \ \tilde{B}.
\end{gather}
Now define two functions $\Psi_1:[0,+\infty)\longrightarrow[0,1]$,
$\Psi_2:[0,+\infty)\longrightarrow[0,p(H)]$ by
\begin{align}\label{definicja Psi_1}
\Psi_1(c)&:=P(A_c),\\[2ex]\label{definicja Psi_2}
\Psi_2(c)&:=P^{\ast}(A_c)\cdot
\mathbf{\tilde{E}}[e^{-rT}H]=\mathbf{\tilde{E}}[e^{-rT}H\mathbf{1}_{A_c}].
\end{align}
Let us notice that both functions $\Psi_1, \Psi_2$ are decreasing
and $\Psi_1(0)=1$, $\Psi_2(0)=p(H)$. Thus $\Psi_2(0)$ provides the
arbitrage free price of the continent claim $H$. Below we list
some properties of functions $\Psi_1,\Psi_2$ needed in the sequel.
First let us introduce two conditions concerning the real function
$f:\mathbb{R}^d\longrightarrow[0,+\infty)$: 
\begin{enumerate}[(C1)]
\item $\lambda_d\big(\{z: f(z)=c\}\big)=0$ for each $c>0$,
\item $\lambda_d\big(\{z:f(z)\in(a,b]\}\big)>0$ for each $0<a<b$.
\end{enumerate}
Above $\lambda_d$ stands for the Lebesgue measure on $\mathbb{R}^d$.
\begin{prop}\label{proposition o Psi_1}
\begin{enumerate}[a)]
\item The function $\Psi_1$ is left continuous with right hand side limits in each point
of the domain.
\item The following holds
\begin{gather*}
\lim_{c\rightarrow+\infty}\Psi_1(c)=P(H=0).
\end{gather*}
\noindent Assume that $\tilde{Z}_TH=f(W_T)$ where
$f:\mathbb{R}^d\longrightarrow[0,+\infty)$. Then $\Psi_1$ is
\item continuous if and only if $(C1)$ is satisfied,
\item strictly decreasing if and only if $(C2)$ is satisfied.
\end{enumerate}
\end{prop}
{\bf Proof:} a) The function $\Psi_1$ can be written in the form
\begin{gather}\label{postac Psi_1 z dystrybuanta}
\Psi_1(c)=P\left(\tilde{Z}_TH\leq\frac{1}{c}\right)=F_{\tilde{Z}_TH}\left(\frac{1}{c}\right),
\quad c>0,
\end{gather}
where $F_{\tilde{Z}_TH}$ stands for the distribution function of the
random variable $\tilde{Z}_TH$. Thus $\Psi_1$ has one sided limits
for any $c>0$ and the left continuity follows from the right
continuity of $F_{\tilde{Z}_TH}$ for any $c>0$. Left continuity at
$c=0$ follows from monotonicity.\\
$b)$ The assertion follows from the formula
\begin{gather*}
\Psi_1(c)=P(\tilde{Z}^{-1}_T\geq cH\mid
H>0)P(H>0)+P(\tilde{Z}^{-1}_T\geq cH\mid H=0)P(H=0)
\end{gather*}
and the following
\begin{gather*}
\lim_{c\rightarrow +\infty}P(\tilde{Z}^{-1}_T\geq cH\mid H>0)=0.
\end{gather*}
$c)$ First show continuity at zero. If $c_n\downarrow0$ then
$\{\tilde{Z}^{-1}_T\geq c_nH\}_n$ is an increasing family of sets
and by the continuity of probability we have
\begin{align*}
\lim_{n\rightarrow+\infty}\Psi_1(c_n)&=\lim_{n\rightarrow
+\infty}P(\tilde{Z}^{-1}_T\geq
c_nH)=P\Big(\bigcup_{n}\{\tilde{Z}^{-1}_T\geq
c_nH\}\Big)\\[1ex]&=P(\tilde{Z}^{-1}_T>0)=1=\Psi_1(0).
\end{align*}
Taking into account \eqref{postac Psi_1 z dystrybuanta} we see that
$\Psi_1$ is continuous for each $c>0$ if and only if the random
variable $\tilde{Z}_TH=f(W_T)$ has no positive atoms. In view of the
equality
\begin{gather*}
P(\tilde{Z}_TH=c)=P(f(W_T)=c)=\mathcal{L}_{W_T}\Big(\{z:
f(z)=c\}\Big), \quad c>0,
\end{gather*}
and the fact that the distribution of $W_T$ is nondegenerate we see
that the continuity of $\Psi_1$ is equivalent to $(C1)$. $\mathcal{L}_{W_T}$ above stands for the
distribution of $W_T$.\\
$d)$ For $0<c_1<c_2$ we have
\begin{align*}
\Psi_1(c_1)-\Psi_1(c_2)&=P\Big(\tilde{Z}_TH\leq\frac{1}{c_1}\Big)-P\Big(\tilde{Z}_TH\leq\frac{1}{c_2}\Big)\\[1ex]
&=P\Big(f(W_T)\in\Big(\frac{1}{c_2},\frac{1}{c_1}\Big]\Big)=\mathcal{L}_{W_T}\Big(\Big\{z:
f(z)\in\Big(\frac{1}{c_2},\frac{1}{c_1}\Big]\Big\}\Big),
\end{align*}
and it follows from the nondegeneracy of the distribution of $W_T$
that the strict monotonicity of $\Psi_1$ is equivalent to $(C_2)$.
\hfill$\square$

\begin{prop}\label{proposition o Psi_2}
\begin{enumerate}[a)]
\item The function $\Psi_2$ is left continuous with right hand side limits in each point
of the domain.
\item The following holds
\begin{gather*}
\lim_{c\rightarrow+\infty}\Psi_2(c)=0.
\end{gather*}
\noindent Assume that $\tilde{Z}_TH=f(W_T)$ where
$f:\mathbb{R}^d\longrightarrow[0,+\infty)$. Then $\Psi_2$ is
\item continuous if and only if $(C1)$ is satisfied,
\item strictly decreasing if and only if $(C2)$ is satisfied.
\end{enumerate}
\end{prop}
{\bf Proof:} $a)$ It follows from monotonicity that one
sided limits exist. We show left continuity for any $c>0$. For
$c_n\uparrow c$ the family
\begin{gather*}
\{\tilde{Z}^{-1}_T\geq c_nH\}_n
\end{gather*}
is decreasing and
\begin{gather*}
\bigcap_n\{\tilde{Z}^{-1}_T\geq
c_nH\}=\{H=0\}\cup\{\tilde{Z}^{-1}_T\geq cH\}=\{\tilde{Z}^{-1}_T\geq
cH\}.
\end{gather*}
Thus by the dominated convergence we have
\begin{gather*}
\lim_{n\rightarrow+\infty}\Psi_2(c_n)=\lim_{n\rightarrow+\infty}\mathbf{\tilde{E}}[e^{-rT}H\mathbf{1}_{\{\tilde{Z}^{-1}_T\geq
c_nH\}}]=\mathbf{\tilde{E}}[e^{-rT}H\mathbf{1}_{\{\tilde{Z}^{-1}_T\geq
cH\}}]=\Psi_2(c).
\end{gather*}
$b)$ For $c_n\uparrow +\infty$ we have
\begin{gather*}
\{\tilde{Z}^{-1}_T\geq
c_nH\}_n\downarrow\bigcap_n\{\tilde{Z}^{-1}_T\geq
c_nH\}=\{H=0\}\cup\{\tilde{Z}^{-1}_T=+\infty\}=\{H=0\},
\end{gather*}
and thus
\begin{gather*}
\lim_{n\rightarrow+\infty}\Psi_2(c_n)=\lim_{n\rightarrow+\infty}\mathbf{\tilde{E}}[e^{-rT}H\mathbf{1}_{\{\tilde{Z}^{-1}_T\geq
c_nH\}}]=\mathbf{\tilde{E}}[e^{-rT}H\mathbf{1}_{\{H=0\}}]=0.
\end{gather*}
$c)$ We show that the right continuity of $\Psi_2$ is equivalent to
$(C1)$. Then continuity follows from $(a)$. For $c_n\downarrow
c\geq0$ we have
\begin{align*}
\{\tilde{Z}^{-1}_T\geq
c_nH\}\uparrow\bigcup_{n}\{\tilde{Z}^{-1}_T\geq
c_nH\}&=\{H=0\}\cup\{H>0,\tilde{Z}^{-1}_T>cH\}\\[2ex]
&=\{\tilde{Z}^{-1}_T>cH\}=\{1>cf(W_T)\},
\end{align*}
and thus
\begin{gather*}
\lim_{n\rightarrow+\infty}\Psi_2(c_n)={\mathbf{\tilde{E}}}[e^{-rT}H\mathbf{1}_{\{1>cf(W_T)\}}].
\end{gather*}
The condition $\lim_{n\rightarrow+\infty}\Psi_2(c)=\Psi_2(c)$ holds
if and only if $\tilde{P}(1\geq cf(W_T))=\tilde{P}(1>cf(W_T))$. The
last condition holds for $c=0$ and for $c>0$ it is equivalent to
$(C1)$.\\
$d)$ Fix $0<c_1<c_2$. The inequality
\begin{gather*}
\Psi_2(c_1)-\Psi_2(c_2)=\mathbf{\tilde{E}}[e^{-rT}H\mathbf{1}_{\{\frac{1}{c_1}<f(W_T)\leq\frac{1}{c_2}\}}]>0
\end{gather*}
holds if and only if
$\tilde{P}(\frac{1}{c_1}<f(W_T)\leq\frac{1}{c_2})>0$. The last
condition is equivalent to $(C2)$.\hfill$\square$

\vskip2ex

Now assume that $\tilde{Z}_TH=f(W_T)$ for some
$f:\mathbb{R}^{d}\longrightarrow[0,+\infty)$. Let us fix
$\alpha\in[0,1]$, $x>0$ and consider the problem of existence of
solutions to the equation
\begin{gather}\label{rownanie z Psi_1}
\Psi_1(c)=1-\alpha,
\end{gather}
as well as to
\begin{gather}\label{rownanie z Psi_2}
\Psi_2(c)=x.
\end{gather}
In view of Propositions \ref{proposition o Psi_1} and
\ref{proposition o Psi_2} it follows that if $(C1)$ is satisfied
then $\Psi_1$, $\Psi_2$ are continuous decreasing functions with
images $(P(H=0),1]$ and $(0,p(H)]$ respectively. Thus for
$\alpha\in[0, P(H\neq0))$ and $x\in(0,p(H)]$ the equations
\eqref{rownanie z Psi_1} and \eqref{rownanie z Psi_2} do have
solutions. Moreover, if $(C2)$ is satisfied then the solutions are
unique.

The description of functions $\Phi_1$ and $\Phi_2$ is provided by
the following theorem, which is the main result of the paper.
\begin{tw}\label{TH MAIN}
Assume that $\tilde{Z}_TH=f(W_T)$ for some
$f:\mathbb{R}^{d}\longrightarrow[0,+\infty)$ satisfying $(C1)$.
\begin{enumerate}[a)]
\item Let $c=c(x)\in[0,+\infty)$ be a solution of the equation
\begin{gather}\label{warunek na Psi_2}
\Psi_2(c)=x, \qquad x\in(0,p(H)).
\end{gather}
Then the maximal probability function is given by
\begin{gather*}
\Phi_1(x)=
\begin{cases}
P(H=0) \ &\text{for} \  x=0,\\
\Psi_1(c(x)) \ &\text{for}\ x\in(0,p(H)),\\
1 &\text{for} \ x\geq p(H)
\end{cases}
\end{gather*}
Moreover, for any $x\in(0,p(H))$ the probability maximizing strategy
for $x$ is that one replicating the payoff $H\mathbf{1}_{A_{c(x)}}$.
\item Let $c=c(\alpha)\in[0,+\infty)$ be a solution of the equation
\begin{gather}\label{warunek na Psi_1}
\Psi_1(c)=1-\alpha, \qquad \alpha\in [0,P(H\neq0)).
\end{gather}
Then the cost reduction function is given by
\begin{gather*}
\Phi_2(\alpha)=
\begin{cases}
\Psi_2(c(\alpha))&\text{for} \ \alpha\in[0,P(H\neq0)),\\
0 &\text{for} \ \alpha\in[P(H\neq0),1].
\end{cases}
\end{gather*}
Moreover, for any $\alpha\in[0,P(H\neq0))$ the cost reduction
strategy for $\alpha$ is that one replicating the payoff
$H\mathbf{1}_{A_{c(\alpha)}}$.
\end{enumerate}
\end{tw}
{\bf Proof:} The proof is based on the consideration proceeding the
formulation of the Theorem.

$a)$ If $x\geq p(H)$ then the hedging strategy is the probability
maximizing strategy and then clearly $\Phi_1(x)=1$. Consider the
case $x\in(0,p(H))$. By Theorem \ref{TH 1 max prob} we know that
$\Phi_1(x)=P(\tilde A)$, where $\tilde{A}$ is a solution of
\eqref{twierdzenie o tilde A}. The solution of \eqref{TH 1 max prob
(2)}, which is equivalent to \eqref{twierdzenie o tilde A}, is of
the form \eqref{postac A_c} with $c$ satisfying \eqref{warunek na
c(x) (2)}. But \eqref{warunek na c(x) (2)} is equivalent to
\eqref{warunek na Psi_2}. Thus we have
\begin{gather*}
\Phi_1(x)=P(A_c)=\Psi_1(c),
\end{gather*}
where $c$ is given by the condition $\Psi_2(c)=x$. For $x=0$
consider the trivial strategy $\pi=0$. Then $P(X_T^{x,\pi}\geq
H)=P(H=0)$. On the other hand, due to the monotonicity of $\Phi_1$,
we have $\Phi_1(0)\leq\lim_{x\downarrow
0}\Phi_1(x)=\lim_{x\downarrow
0}\Psi_1(c(x))=\lim_{z\uparrow+\infty}\Psi_1(z)=P(H=0)$. As a
consequence we obtain $\Phi_1(0)=P(H=0)$. The second part of the
assertion follows from Theorem \ref{TH 1 max prob}.

$b)$ If $\alpha\in[P(H\neq0),1]$ then consider a trivial strategy
$\pi=0$ with zero initial endowment $x=0$. Then $X_T^{x,\pi}=0$ and
thus $P(X_T^{x,\pi}\geq H)=P(H=0)\geq1-\alpha$. As a consequence we
have $\Phi_2(\alpha)=0$. Now consider the case $\alpha\in[0,P(H\neq
0)]$. It follows from Theorem \ref{TH 2 min cost} that
$\Phi_2(\alpha)=\mathbf{\tilde{E}}[e^{-rT}H\mathbf{1}_{\tilde{B}}]$,
where $\tilde{B}$ is a solution to \eqref{twierdzenie o tilde B}.
The optimal solution of \eqref{twierdzenie o tilde B} is the same as
for \eqref{TH 2 min cost (2)} and has the form \eqref{postac A_c}
with $c$ satisfying \eqref{warunek na c(alfa) (2)}. The condition
\eqref{warunek na c(alfa) (2)} can be written as
$\Psi_1(c)=1-\alpha$. Thus we have
\begin{gather*}
\Phi_2(\alpha)=\mathbf{\tilde{E}}[e^{-rT}H\mathbf{1}_{A_c}]=\Psi_2(c).
\end{gather*}
The second part of the assertion follows from Theorem \ref{TH 2 min
cost}. \hfill $\square$

\vskip2ex

In virtue of Theorem \ref{TH MAIN} the only problem to determine
functions $\Phi_1$, $\Phi_2$  is to find functions $\Psi_1$,
$\Psi_2$ and to solve the equations \eqref{warunek na Psi_2},
\eqref{warunek na Psi_1}. In general, due to the fact that $\Psi_1,
\Psi_2$ are rather of a sophisticated form, one should not expect to
find analytic formulas for the constants in \eqref{warunek na
Psi_2}, \eqref{warunek na Psi_1}. However, the equations
\eqref{warunek na Psi_2}, \eqref{warunek na Psi_1} can be solved
with the use of numerical methods. In the sequel we solve the
problem of determining functions $\Psi_1$, $\Psi_2$ for the most
common basket derivatives.

\section{Quantile hedging in two dimensional model}\label{Quantile hedging in two dimensional model}
In this section we determine explicit formulas for the functions
$\Psi_1$, $\Psi_2$ for a few examples of popular options. Since our
derivatives depend on two underlying assets we simplify at the
beginning general formulas from Section \ref{Description of the
method} to the case $d=2$. In the calculations we base on properties
of the multidimensional normal distribution which are recalled in
the sequel.

For the case $d=2$ we denote the correlation matrix by
\begin{gather*}
Q=\left[\begin{array}{rr}
1&\rho\\
\rho&1\\
\end{array}\right].
\end{gather*}
Consequently, we have
\begin{gather*}
Q^{-1}=\frac{1}{\rho^2-1}\left[\begin{array}{rr}
-1&\rho\\
\rho&-1\\
\end{array}\right],\qquad
Q^{-\frac{1}{2}}=\frac{1}{2}\left[\begin{array}{rr}
\frac{1}{\sqrt{1+\rho}}+\frac{1}{\sqrt{1-\rho}}&\frac{1}{\sqrt{1+\rho}}-\frac{1}{\sqrt{1-\rho}}\\[2ex]
\frac{1}{\sqrt{1+\rho}}-\frac{1}{\sqrt{1-\rho}}&\frac{1}{\sqrt{1+\rho}}+\frac{1}{\sqrt{1-\rho}}\\
\end{array}\right].
\end{gather*}
Hence the density of the martingale measure \eqref{gestosc miary
martyngalowej - ogolnie} can be written as
\begin{gather}\label{wzor koncowy na tildeZ_T}
\tilde{Z}_T=e^{-A_1W^1_T-A_2W^2_T-BT},
\end{gather}
where
\begin{align*}
A_1&:=\frac{1}{\rho^2-1}\left(-\frac{\alpha_1-r}{\sigma_1}+\rho\frac{\alpha_2-r}{\sigma_2}\right), \quad A_2:=\frac{1}{\rho^2-1}\left(\rho\frac{\alpha_1-r}{\sigma_1}-\frac{\alpha_2-r}{\sigma_2}\right),\\[1ex]
B&:=\frac{1}{8}\Bigg(\bigg(\Big(\frac{1}{\sqrt{1+\rho}}+\frac{1}{\sqrt{1-\rho}}\Big)\frac{\alpha_1-r}{\sigma_1}
+\Big(\frac{1}{\sqrt{1+\rho}}-\frac{1}{\sqrt{1-\rho}}\Big)\frac{\alpha_2-r}{\sigma_2}\bigg)^2\\[1ex]
&+\bigg(\Big(\frac{1}{\sqrt{1+\rho}}-\frac{1}{\sqrt{1-\rho}}\Big)\frac{\alpha_1-r}{\sigma_1}
+\Big(\frac{1}{\sqrt{1+\rho}}+\frac{1}{\sqrt{1-\rho}}\Big)\frac{\alpha_2-r}{\sigma_2}\bigg)^2
\Bigg).
\end{align*}
The formula \eqref{postac A_c} for the set $A_c$ simplifies to the
form
\begin{gather*}
A_c =\left\{\tilde{Z}_T^{-1}\geq
cH\right\}=\left\{e^{A_1W^1_T+A_2W^2_T+BT}\geq c H\right\},
\end{gather*}
and consequently formulas \eqref{definicja Psi_1}, \eqref{definicja
Psi_2} become
\begin{align*}
\Psi_1(c)&=P(e^{A_1W^1_T+A_2W^2_T+BT}\geq c H),\\[2ex]
\Psi_2(c)&=\mathbf{\tilde{E}}[e^{-rT}H\mathbf{1}_{\{e^{A_1W^1_T+A_2W^2_T+BT}\geq
c H\}}].
\end{align*}

Now set the notation concerning the multidimensional normal
distribution and recall its basic properties, which can be found in
standard textbooks on probability theory or statistics, see for
instance \cite{Joh-Wich}. A random vector $X$ taking values in
$\mathbb{R}^d$ has a multidimensional normal distribution if its
density is of the form
\begin{gather}\label{gestosc rozkladu nor}
f_X(x)=\frac{1}{(2\pi)^\frac{d}{2}(\text{det}
\Sigma)^{\frac{1}{2}}}\cdot e^{-\frac{1}{2}(x-m)^T\Sigma^{-1}(x-m)},
\qquad x\in\mathbb{R}^d,
\end{gather}
where $m\in\mathbb{R}^d$ is a mean of $X$ and $\Sigma$ is a
symmetric positive definite $d\times d$ covariance matrix of $X$.
The fact that $X$ has a density \eqref{gestosc rozkladu nor} will be
denoted by $X\sim N_d(m,\Sigma)$ or $\mathcal{L}(X)=N_d(m,\Sigma)$.
If $d=1$ then the subscript is omitted and $N(m,\sigma)$ denotes the
normal distribution with mean $m$ and variance $\sigma$. If $X\sim
N_d(m,\Sigma)$ and $A$ is a $k\times d$ matrix then,
\begin{gather}\label{rozklad AX}
AX\sim N_k(Am,A\Sigma A^T);
\end{gather}
in particular if $a\in\mathbb{R}^{d}$ then
\begin{gather}\label{rozklad aX}
a^{T}X\sim N(a^Tm,a^T\Sigma a).
\end{gather}
Let $X$ be a random vector taking values in $\mathbb{R}^d$ and fix
an integer $0<k<d$. Let us divide $X$ into two vectors $X^{(1)}$ and
$X^{(2)}$ with lengths $k$, $d-k$ respectively, i.e.
\begin{gather*}
X^{(1)}=(X_1,X_2,...,X_k)^T, \qquad
X^{(2)}=(X_{k+1},X_{k+2},...,X_d)^T.
\end{gather*}
Analogously, divide the mean vector $m$ and the covariance matrix
$\Sigma$
\begin{displaymath}
m= \left(
\begin{array}{c}
m^{(1)}\\[1ex]
m^{(2)}
\end{array}
\right); \qquad \Sigma= \left[
\begin{array}{cc}
\Sigma^{(11)}&\Sigma^{(12)}\\[1ex]
\Sigma^{(21)}&\Sigma^{(22)}
\end{array}
\right],
\end{displaymath}
so that $\mathbf{E}X^{(1)}=m^{(1)}$, $\mathbf{E}X^{(2)}=m^{(2)}$,
$Cov X^{(1)}=\Sigma^{(11)}$, $Cov X^{(2)}=\Sigma^{(22)}$, $Cov
(X^{(1)},X^{(2)})=\Sigma^{(12)}={\Sigma^{(21)}}^T$. Denote by
$\mathcal{L}\left(X^{(1)}\mid X^{(2)}=x^{(2)}\right)$ the
conditional distribution of $X^{(1)}$ given
$X^{(2)}=x^{(2)}\in\mathbb{R}^{d-k}$. If $\Sigma^{(22)}$ is
nonsingular then
\begin{gather}\label{tw o rozkl nor warunkowym}
\mathcal{L}\left(X^{(1)}\mid
X^{(2)}=x^{(2)}\right)=N_{k}(m^{(1)}(x^{(2)}),\Sigma^{(11)}(x^{(2)})),
\end{gather}
where
\begin{align}\label{wzor na srednia i wariancje waunkowa ogolny}\nonumber
m^{(1)}(x^{(2)})&=
m^{(1)}+\Sigma^{(12)}{\Sigma^{(22)}}^{-1}(x^{(2)}-m^{(2)}), \\[2ex]
\Sigma^{(11)}(x^{(2)})&=\Sigma^{(11)}-\Sigma^{(12)}{\Sigma^{(22)}}^{-1}\Sigma^{(21)}.
\end{align}
Actually the conditional variance $\Sigma^{(11)}(x^{(2)})$ does not
depend on $x^{(2)}$ but we keep the notation for the sake of
consistency. The conditional density will be denoted by
$f_{X^{(1)}\mid X^{(2)}=x^{(2)}}(x^{(1)})$, where
$x^{(1)}\in\mathbb{R}^k$. In particular if $(X,Y)$ is a two
dimensional normal vector with parameters
\begin{displaymath}
m= \left(
\begin{array}{c}
m_1\\[1ex]
m_2
\end{array}
\right); \qquad \Sigma= \left[
\begin{array}{cc}
\sigma_{11}&\sigma_{12}\\[1ex]
\sigma_{21}&\sigma_{22}
\end{array}
\right],
\end{displaymath}
then
\begin{gather*}
\mathcal{L}(X\mid Y=y)=N(m_1(y),\sigma_1(y)),
\end{gather*}
where
\begin{gather}\label{wzor na sredia i wariancja warunkowa szczegolny}
m_1(y):= m_1+\frac{\sigma_{12}}{\sigma_{22}}(y-m_2), \quad
\sigma_1(y):= \sigma_{11}-\frac{\sigma_{12}^2}{\sigma_{22}}.
\end{gather}
If $X$ is a random vector then its distribution wrt. the measure
$\widetilde{P}$ will be denoted by $\tilde{\mathcal{L}}(X)$ and its
density by $\tilde{f}_X$. Analogously, $\tilde{f}_{X^{(1)}\mid
X^{(2)}=x^{(2)}}(x^{(1)})$ stands for the conditional density with
respect to the measure $\widetilde{P}$.

In the following subsections we will use the universal constants:
$A_1, A_2, B$ defined in \eqref{wzor koncowy na tildeZ_T} as well as
$a_1, a_2, b, \tilde{a}_1, \tilde{a}_2,
\tilde{b}$ introduced below.\\
\noindent Fix a number $K>0$. One can check the following
\begin{align}\label{wzor S^1>K}
\left\{S^1_T\geq K\right\}&=\left\{W^1_T\geq a_1\right\}=\left\{\widetilde{W}^1_T\geq \tilde{a}_1\right\},\\[2ex]\label{wzor S^2>K}
\left\{S^2_T\geq K\right\}&=\left\{W^2_T\geq a_2\right\}=\left\{\widetilde{W}^2_T\geq \tilde{a}_2\right\},\\[2ex]\label{wzor S^1>S^2}
\left\{S^1_T\geq
S^2_T\right\}&=\left\{\sigma_1W^1_T-\sigma_2W^2_T\geq
b\right\}=\left\{\sigma_1\widetilde{W}^1_T-\sigma_2\widetilde{W}^2_T\geq
\tilde{b}\right\},
\end{align}
where{\small
\begin{align*}
a_1&:=\frac{1}{\sigma_1}\left(\ln\frac{K}{S^1_0}-(\alpha_1-\frac{1}{2}\sigma_1^2)T\right),\quad
\tilde{a}_1:=\frac{1}{\sigma_1}\left(\ln\frac{K}{S^1_0}-(r-\frac{1}{2}\sigma_1^2)T\right),
\\[2ex]
a_2&:=\frac{1}{\sigma_2}\left(\ln\frac{K}{S^2_0}-(\alpha_2-\frac{1}{2}\sigma_2^2)T\right),\quad
\tilde{a}_2:=\frac{1}{\sigma_2}\left(\ln\frac{K}{S^2_0}-(r-\frac{1}{2}\sigma_2^2)T\right)\\[2ex]
b&:=\ln\left(\frac{S^2_0}{S^1_0}\right)
+(\alpha_2-\alpha_1-\frac{1}{2}(\sigma_2^2-\sigma_1^2))T, \quad
\tilde{b}:=\ln\left(\frac{S^2_0}{S^1_0}\right)
-\frac{1}{2}(\sigma_2^2-\sigma_1^2)T.
\end{align*}}
In all the formulas appearing in the sequel it is understood that
$\ln0=-\infty$ and $\Phi$ stands for the cumulative distribution function of $N(0,1)$.

\subsection{Digital option}
In this section we determine $\Psi_1, \Psi_2$ for the payoff
\begin{gather}\label{digital option}
H=K \cdot \mathbf{1}_{\{S^1_T\geq S^2_T\}},\quad \text{where} \quad
K>0.
\end{gather}

\noindent By \eqref{wzor S^1>S^2} we have 

\begin{align}\label{P(A_c) digital option}\nonumber
\Psi_1(c)&=P(A_c)=P\left(e^{A_1W^1_T+A_2W^2_T+BT}\geq cK
\mathbf{1}_{\{S^1_T\geq
S^2_T\}}\right)=P(A_1W^1_T+A_2W^2_T\\[1ex]\nonumber
&\phantom{=}+BT\geq \ln (cK), S^1_T\geq S^2_T)+P(e^{A_1W^1_T+A_2W^2_T+BT}\geq 0,
S^1_T< S^2_T)\\[1ex]\nonumber
&=P\left(A_1W^1_T+A_2W^2_T+BT\geq\ln (cK)\mid
\sigma_1W^1_T-\sigma_2W^2_T\geq b
\right)\\[1ex]
&\phantom{=}\cdot P(\sigma_1W^1_T-\sigma_2W^2_T\geq b)+P(\sigma_1W^1_T-\sigma_2W^2_T< b).
\end{align}
\noindent
Let us notice that
\begin{gather*}
X:= \left[\begin{array}{r}
A_1W^1_T+A_2W^2_T\\
\sigma_1W^1_T-\sigma_2W^2_T
\end{array}\right]
=
\left[\begin{array}{rr}
A_1&A_2\\
\sigma_1&-\sigma_2\\
\end{array}\right]
 \left[\begin{array}{r}
W^1_T\\
W^2_T\\
\end{array}\right],
\end{gather*}
so in view of \eqref{rozklad AX} we have $X\sim N_2(0,\Sigma)$,
where
\begin{gather*}
\Sigma=\left[\begin{array}{rr}
(A_1+A_2r)TA_1+(A_1r+A_2)TA_2&(\sigma_1-\sigma_2r)TA_1+(\sigma_1r-\sigma_2)TA_2\\[1ex]
(\sigma_1-\sigma_2r)TA_1+(\sigma_1r-\sigma_2)TA_2&(\sigma_1-\sigma_2r)T\sigma_1-(\sigma_1r-\sigma_2)T\sigma_2
\end{array}\right].
\end{gather*}
In virtue of \eqref{wzor na sredia i wariancja warunkowa szczegolny}
we have
\begin{gather*}
\mathcal{L}(A_1W^1_T+A_2W^2_T\mid\sigma_1W^1_T-\sigma_2W^2_T=y)=N(m(y),\sigma(y)),
\end{gather*}
where
\begin{gather*}
m(y)=y\frac{(\sigma_1-\sigma_2r)A_1+(\sigma_1r-\sigma_2)A_2}{(\sigma_1-\sigma_2r)\sigma_1-(\sigma_1r-\sigma_2)\sigma_2};\qquad
\sigma(y)=\frac{T(A_1\sigma_2+A_2\sigma_1)^2(\rho^2-1)}{-\sigma_1^2+2\rho\sigma_1\sigma_2-\sigma_2^2}.
\end{gather*}
By \eqref{rozklad aX} we have: $\sigma_1W^1_T-\sigma_2W^2_T\sim
N(0,T(\sigma_1^2-2\rho\sigma_1\sigma_2+\sigma_2^2))$. Going back to
\eqref{P(A_c) digital option} we have 
\begin{align*}
&\Psi_1(c)=\int_{b}^{+\infty}P(A_1W^1_T+A_2W^2_T\geq\ln (cK)-BT\mid
\sigma_1W^1_T-\sigma_2W^2_T=y)\\[1ex]&\cdot f_{\sigma_1W^1_T-\sigma_2W^2_T}(y)dy+P(\sigma_1W^1_T-\sigma_2W^2_T< b)=\hskip-1ex\int_{b}^{+\infty}\Phi\left(\frac{m(y)+BT-\ln
(cK)}{\sqrt{\sigma(y)}}\right)\\[1ex]
&\cdot f_{\sigma_1W^1_T-\sigma_2W^2_T}(y)dy+\Phi\left(\frac{b}{\sqrt{T(\sigma_1^2-2\rho\sigma_1\sigma_2+\sigma_2^2)}}\right).
\end{align*}

\noindent Now let us determine $\Psi_2$. In virtue of \eqref{wzor
S^1>S^2} we have {\small
\begin{align*}
&\Psi_2(c)=e^{-rT}\mathbf{\tilde{E}}[H\mathbf{1}_{A_c}]=e^{-rT}\mathbf{\tilde{E}}[K\mathbf{1}_{\{S^1_T\geq
S^2_T\}}\cdot\mathbf{1}_{\{\tilde{Z}^{-1}_T\geq
cK\mathbf{1}_{\{S^1_T\geq
S^2_T\}}\}}]\\[1ex]&=e^{-rT}K\widetilde{P}\left(S^1_T\geq
S^2_T,\tilde{Z}^{-1}_T\geq cK\mathbf{1}_{\{S^1_T\geq
S^2_T\}}\right)\\[1ex]
&=e^{-rT}K\widetilde{P}\left(\tilde{Z}^{-1}_T\geq cK\mid S^1_T\geq
S^2_T\right)\widetilde{P}(S^1_T\geq S^2_T)\\[1ex]
&=e^{-rT}K\widetilde{P}(e^{A_1W^1_T+A_2W^2_T+BT}>cK\mid
\sigma_1\widetilde{W}^1_T-\sigma_2\widetilde{W}^2_T\geq
\tilde{b})\widetilde{P}(\sigma_1\widetilde{W}^1_T-\sigma_2\widetilde{W}^2_T\geq \tilde{b})\\[1ex]
&=e^{-rT}K\int_{\tilde{b}}^{+\infty}\widetilde{P}(e^{A_1W^1_T+A_2W^2_T+BT}>cK\mid
\sigma_1\widetilde{W}^1_T-\sigma_2\widetilde{W}^2_T=y)\tilde{f}_{\sigma_1\widetilde{W}^1_T-\sigma_2\widetilde{W}^2_T}(y)dy
\\[1ex]
&=e^{-rT}K
\cdot\int_{\tilde{b}}^{+\infty}\widetilde{P}(A_1\widetilde{W}^1_T+A_2\widetilde{W}^2_T>\ln
(cK)+A_1\frac{\alpha_1-r}{\sigma_1}T+A_2\frac{\alpha_2-r}{\sigma_2}T\\[1ex]
&\phantom{=}-BT\mid
\sigma_1\widetilde{W}^1_T-\sigma_2\widetilde{W}^2_T=y)\cdot\tilde{f}_{\sigma_1\widetilde{W}^1_T-\sigma_2\widetilde{W}^2_T}(y)dy
=e^{-rT}K\\[1ex]
&\phantom{=}\cdot \int_{\tilde{b}}^{+\infty}\Phi\left(\frac{m(y)-\ln
(cK)-A_1\frac{\alpha_1-r}{\sigma_1}T-A_2\frac{\alpha_2-r}{\sigma_2}T+BT}{\sqrt{\sigma(y)}}\right)\tilde{f}_{\sigma_1\widetilde{W}^1_T-\sigma_2\widetilde{W}^2_T}(y)dy.
\end{align*}
}

\subsection{Quantos}
\subsubsection{Quanto domestic}
The contingent claim is of the form
\begin{gather}\label{quanto domestic}
H=S^2_T(S^1_T-K)^{+}, \quad K>0.
\end{gather}
At the beginning let us notice that {\small
\begin{align}\label{wzor 1 QD}\nonumber
A_c=\left\{e^{A_1W^1_T+A_2W^2_T+BT}\geq c S^2_T(S^1_T-K)\right\}&=
\left\{(A_2-\sigma_2)W^2_T\geq v(c,W^1_T)\right\}\\[1ex]
&= \left\{(A_2-\sigma_2)\widetilde{W}^2_T\geq w(c,\widetilde{W}^1_T)\right\},
\end{align}}
where
\begin{align*}
v(c,x)&:=\ln\left(cS^2_0e^{(\alpha_2-\frac{1}{2}\sigma_2^2-B)T-A_1x}(S_0^1e^{(\alpha_1-\frac{1}{2}\sigma_1^2)T+\sigma_1x}-K)\right),\\[2ex]
w(c,x)&:=\ln\left[cS^2_0e^{(r-\frac{1}{2}\sigma_2^2-B+A_1\frac{\alpha_1-r}{\sigma_1}+A_2\frac{\alpha_2-r}{\sigma_2})T-A_1x}(S_0^1e^{(r-\frac{1}{2}\sigma_1^2)T+\sigma_1x)}-K)\right].
\end{align*}
By \eqref{wzor S^1>K} and \eqref{wzor 1 QD} we have 
\begin{align*} &\Psi_1(c)=P\left(A_c\mid
S^1_T\geq K\right)P(S^1_T\geq K)+P\left(A_c\mid S^1_T<
K\right)P(S^1_T<K)\\[1ex]
&=P\left((A_2-\sigma_2)W^2_T\geq
\ln\left(cS^2_0e^{(\alpha_2-\frac{1}{2}\sigma_2^2-B)T-A_1W^1_T}(S^1_T-K)\right)\mid
W^1_T\geq a_1\right)\\[1ex]
&\phantom{=}\cdot P(W^1_T\geq a_1)+P(W^1_T< a_1)\\[1ex]
&=\int_{a_1}^{+\infty}P((A_2-\sigma_2)W^2_T\geq v(c,W^1_T)\mid
W^1_T=x)f_{W^1_T}(x)dx+\Phi\left(\frac{a_1}{\sqrt{T}}\right).
\end{align*}
The conditional distribution is given by
\begin{gather*}
\mathcal{L}((A_2-\sigma_2)W^2_T\mid W^1_T=x)\sim N(m(x),\sigma(x)),
\end{gather*}
where $m(x),\sigma(x)$ are given by \eqref{wzor na srednia i
wariancje waunkowa ogolny}. Hence we have
\begin{align*}
\Psi_1(c)=\int_{a_1}^{+\infty}\Phi\left(\frac{m(x)-v(c,x)}{\sqrt{\sigma(x)}}\right)f_{W^1_T}(x)dx+\Phi\left(\frac{a_1}{\sqrt{T}}\right).
\end{align*}
To avoid technicalities assume that $A_2\neq\sigma_2$. We have
\begin{align*}
&\Psi_2(c)=e^{-rT}\mathbf{\tilde{E}}\left[S^2_T(S^1_T-K)^{+}\mathbf{1}_{A_c}\right]=e^{-rT}\mathbf{\tilde{E}}\left[S^2_T(S^1_T-K)^{+}\mathbf{1}_{A_c}\mid S^1_T\leq K\right]\\[1ex]&\phantom{=}\cdot\widetilde{P}(S^1_T\leq K)+e^{-rT}\mathbf{\tilde{E}}\left[S^2_T(S^1_T-K)^{+}\mathbf{1}_{A_c}\mid S^1_T> K\right]\widetilde{P}(S^1_T> K)\\[2ex]
&=e^{-rT}\mathbf{\tilde{E}}\left[S^2_T(S^1_T-K)\mathbf{1}_{A_c}\mid S^1_T> K\right]\widetilde{P}(S^1_T> K).
\end{align*}
By \eqref{wzor S^1>K} and \eqref{wzor 1 QD} we have 
\begin{align*}
&\Psi_2(c)=e^{-rT}\mathbf{\tilde{E}}\Big[S^2_0e^{(r-\frac{1}{2}\sigma_2^2)T+\sigma_2\widetilde{W}^2_T}(S^1_0e^{(r-\frac{1}{2}\sigma_1^2)T+\sigma_1\widetilde{W}^1_T}-K)\\[1ex]
&\qquad\cdot\mathbf{1}_{\{(A_2-\sigma_2)\widetilde{W}_T^2\geq
w(c,\widetilde{W}^1_T)\}}\mid
\widetilde{W}^1_T>\tilde{a}_1\Big]\widetilde{P}(\widetilde{W}^1_T>\tilde{a}_1)\\[2ex]
&=e^{-rT}\int_{\tilde{a}_1}^{+\infty}\mathbf{\tilde{E}}\Big[S^2_0e^{(r-\frac{1}{2}\sigma_2^2)T+\sigma_2\widetilde{W}^2_T}(S^1_0e^{(r-\frac{1}{2}\sigma_1^2)T+\sigma_1\widetilde{W}^1_T}-K)\\[1ex]&\qquad\cdot\mathbf{1}_{\{(A_2-\sigma_2)\widetilde{W}_T^2\geq
w(c,\widetilde{W}^1_T)\}}\mid
\widetilde{W}^1_T=x\Big]\tilde{f}_{\widetilde{W}^1_T}(x)dx\\[2ex]
&=C_1\int_{\tilde{a}_1}^{+\infty}e^{\sigma_1
x}\int_{w(c,x)}^{+\infty}e^{\frac{\sigma_2}{A_2-\sigma_2}y}\tilde{f}_{(A_2-\sigma_2)\widetilde{W}^2_T\mid\widetilde{W}^1_T=x}(y)dy\tilde{f}_{\widetilde{W}^1_T}(x)dx\\[2ex]
&\phantom{=}-C_2\int_{\tilde{a}_1}^{+\infty}\int_{w(c,x)}^{+\infty}e^{\frac{\sigma_2}{A_2-\sigma_2}y}\tilde{f}_{(A_2-\sigma_2)\widetilde{W}^2_T\mid\widetilde{W}^1_T=x}(y)dy\tilde{f}_{\widetilde{W}^1_T}(x)dx.
\end{align*}
\noindent
with $C_1:=e^{-rT}S_0^1S_0^2e^{(2r-\frac{1}{2}\sigma_1^2-\frac{1}{2}\sigma_2^2)T}, C_2:=e^{-rT}KS^2_0e^{(r-\frac{1}{2}\sigma_2^2)T}$.
It follows from \eqref{wzor na sredia i wariancja warunkowa
szczegolny} that $
\tilde{\mathcal{L}}((A_2-\sigma_2)\widetilde{W}^2_T\mid\widetilde{W}^1_T=x)=N((A_2-\sigma_2)\rho
x,T(1-\rho^2)(A_2-\sigma_2)^2)$ and hence 
{\small
\begin{align*}
\Psi_2(c)
&=C_1 \ \frac{\int_{\tilde{a}_1}^{+\infty}e^{\sigma_1
x}\int_{w(c,x)}^{+\infty}e^{\frac{\sigma_2}{A_2-\sigma_2}y+\frac{(y-(A_2-\sigma_2)\rho x)^2}{2T(1-\rho^2)(A_2-\sigma_2)^2}}dy\tilde{f}_{\widetilde{W}^1_T}(x)dx}{\sqrt{2\pi
T(1-\rho^2)(A_2-\sigma_2)^2}}\\[2ex]
&-C_2\frac{\int_{\tilde{a}_1}^{+\infty}\int_{w(c,x)}^{+\infty}
e^{\frac{\sigma_2}{A_2-\sigma_2}y+\frac{(y-(A_2-\sigma_2)\rho
x)^2}{2T(1-\rho^2)(A_2-\sigma_2)^2}}
dy\tilde{f}_{\widetilde{W}^1_T}(x)dx}{\sqrt{2\pi
T(1-\rho^2)(A_2-\sigma_2)^2}}.
\end{align*}}

\subsubsection{Quanto foreign}
The payoff is of the form
\begin{gather*}
H=\left(S^1_T-\frac{K}{S^2_T}\right)^+, \quad K>0.
\end{gather*}
First let us notice that
\begin{gather}\label{wzor 1 QF}
\left\{S^1_T-\frac{K}{S^2_T}\geq0\right\}=\left\{\sigma_1W^1_T+\sigma_2W^2_T\geq
d\right\}=\left\{\sigma_1\widetilde{W}^1_T+\sigma_2\widetilde{W}^2_T\geq
\tilde{d}\right\}=:\Omega_0,
\end{gather}
where
\begin{align*}
d:=\ln\frac{K}{S^1_0S^2_0}-\left(\alpha_1+\alpha_2-\frac{1}{2}(\sigma_1^2+\sigma_2^2)\right)T,\quad
\tilde{d}:=d+(\alpha_1+\alpha_2-2r)T,
\end{align*}
and
\begin{align}\label{wzor 2 QF}\nonumber
&A_c=\left\{e^{A_1W^1_T+A_2W^2_T+BT}\geq
c\left(S^1_T-\frac{K}{S^2_T}\right)\right\}=\{A_1
W^1_T+(A_2+\sigma_2)W^2_T\\[1ex]&\geq
v(c,\sigma_1W^1_T+\sigma_2W^2_T)\}=\left\{A_1\widetilde{W}^1_T+(A_2+\sigma_2)\widetilde{W}^2_T\geq
w(c,\sigma_1\widetilde{W}^1_T+\sigma_2\widetilde{W}^2_T)\right\},
\end{align}
where
\begin{align*}
v(c,z)&:=\ln\left(\frac{c}{S^2_0}e^{(\frac{1}{2}\sigma_2^2-\alpha_2-B)T}\Big(S^1_0S^2_0e^{\alpha_1+\alpha_2-\frac{1}{2}(\sigma_1^2+\sigma_2^2)T+z}-K\Big)\right),\\[2ex]
w(c,z)&:=\ln\left[\frac{c}{S^2_0}e^{T(\frac{1}{2}\sigma_2^2-r+A_1\frac{\alpha_1-r}{\sigma_1}+A_2\frac{\alpha_2-r}{\sigma_2}-B)}\left(S^1_0S^2_0e^{(2r-\frac{1}{2}(\sigma_1^2+\sigma_2^2))T+z}-K\right)\right].
\end{align*}
By \eqref{wzor 1 QF} we have
\begin{align*}
\Psi_1(c)&=P\left(e^{A_1W^1_T+A_2W^2_T+BT}\geq
c\left(S^1_T-\frac{K}{S^2_T}\right)\mid\Omega_0\right)P(\Omega_0)\\[1ex]
&\phantom{=}+P\left(e^{A_1W^1_T+A_2W^2_T+BT}\geq
0\mid\Omega_0^c\right)P(\Omega_0^c)\\[1ex]
&=P\left(S^2_Te^{A_1W^1_T+A_2W^2_T+BT}\geq
c(S^1_TS^2_T-K)\mid\Omega_0\right)P(\Omega_0)+P(\Omega_0^c),
\end{align*}
As a consequence of \eqref{wzor 2 QF} we obtain
\begin{align*}
&\Psi_1(c) =P(A_1W^1_T+(A_2+\sigma_2)W^2_T\geq
v(c,\sigma_1W^1_T+\sigma_2W^2_T)\mid\Omega_0)P(\Omega_0)\\[1ex]
&\phantom{=}+P(\Omega_0^c)=\int_{d}^{+\infty}P(A_1W^1_T+(A_2+\sigma_2)W^2_T\geq
v(c,z)\mid\sigma_1W^1_T+\sigma_2W^2_T=z)\\[1ex]
&\phantom{=}\cdot f_{\sigma_1W^1_T+\sigma_2W^2_T}(z)dz+P(\Omega_0^c).
\end{align*}
By \eqref{wzor na sredia i wariancja warunkowa szczegolny} we have
\begin{gather*}
\mathcal{L}(A_1W^1_T+(A_2+\sigma_2)W^2_T\mid
\sigma_1W^1_T+\sigma_2W^2_T=z)=N(m(z),\sigma(z)),
\end{gather*}
where
\begin{align*}
m(z)&:=\frac{(A_1+(A_2+\sigma_2)\rho)\sigma_1+(A_1\rho+A_2+\sigma_2)\sigma_2}{\sigma_1^2+2\rho\sigma_1\sigma_2+\sigma_2^2},\\[2ex]
\sigma(z)&:=T\bigg\{\left( A_{1}+\left( A_{2}+\sigma_2\right) \rho
\right) A_{1}+\left( A_{1}\rho+\left( A_{2}+\sigma_2\right)\right)
\left(
A_{2}+\sigma_2\right)\\[2ex]
&-\frac{(\left( A_{1}+\left( A_{2}+\sigma_2\right) \rho \right)
\sigma_1+\left( A_{1}\rho +\left( A_{2}+\sigma_2\right) \right)
\sigma_2)^2}{\sigma_1^2+2\rho\sigma_1\sigma_2+\sigma_2^2}\bigg\},
\end{align*}
and thus
\begin{align*}
\Psi_1(c)&=\int_{d}^{+\infty}\Phi\left(\frac{m(z)-v(c,z)}{\sqrt{\sigma(z)}}\right)f_{\sigma_1W^1_T+\sigma_2W^2_T}(z)dz\\[1ex]
&\phantom{=}+\Phi\left(\frac{d}{\sqrt{T(\sigma_1^2+2\rho\sigma_1\sigma_2+\sigma_2^2})}\right).
\end{align*}
By \eqref{wzor 1 QF} and \eqref{wzor 2 QF} we have
\begin{align*}
&\Psi_2(c)=e^{-rT}\mathbf{\tilde{E}}\left[\left(S^1_T-\frac{K}{S^2_T}\right)\mathbf{1}_{A_c}\mid\Omega_0\right]\widetilde{P}({\Omega}_0)\\[1ex]
&=e^{-rT}\mathbf{\tilde{E}}\left[\left(S^1_T-\frac{K}{S^2_T}\right)\mathbf{1}_{A_c}\mid{\Omega}_0\right]\widetilde{P}({\Omega}_0)\\[2ex]
&=e^{-rT}\int_{\tilde{d}}^{+\infty}\mathbf{\tilde{E}}\left[S^1_T\mathbf{1}_{A_c}\mid
\sigma_1\widetilde{W}^1_T+\sigma_2\widetilde{W}^2_T=z\right]\tilde{f}_{\sigma_1\widetilde{W}^1_T+\sigma_2\widetilde{W}^2_T}(z)dz\\[2ex]
&-e^{-rT}K\int_{\tilde{d}}^{+\infty}\mathbf{\tilde{E}}\left[\frac{1}{S^2_T}\mathbf{1}_{A_c}\mid
\sigma_1\widetilde{W}^1_T+\sigma_2\widetilde{W}^2_T=z\right]\tilde{f}_{\sigma_1\widetilde{W}^1_T+\sigma_2\widetilde{W}^2_T}(z)dz.
\end{align*}
Using \eqref{tw o rozkl nor warunkowym} we find the conditional
distributions
\begin{align*}
\tilde{\mathcal{L}}(\widetilde{W}^1_T,A_1\widetilde{W}^1_T+(A_2+\sigma_2)\widetilde{W}^2_T\mid\sigma_1\widetilde{W}^1_T+\sigma_2\widetilde{W}^2_T=z)&=N_2(M^1(z),\Sigma^1(z)),\\[2ex]
\tilde{\mathcal{L}}(\widetilde{W}^2_T,A_1\widetilde{W}^1_T+(A_2+\sigma_2)\widetilde{W}^2_T\mid\sigma_1\widetilde{W}^1_T+\sigma_2\widetilde{W}^2_T=z)&=N_2(M^2(z),\Sigma^2(z)),
\end{align*}
where $M^1(z), M^2(z), \Sigma^1(z), \Sigma^2(z)$ are determined by
\eqref{wzor na srednia i wariancje waunkowa ogolny}. As a
consequence we obtain
\begin{align*}
&\Psi_2(c)=e^{-rT}S^1_0e^{(r-\frac{1}{2}\sigma_1^2)T}\int_{\tilde{d}}^{+\infty}\int_{-\infty}^{+\infty}\int_{w(c,z)}^{+\infty}e^{\sigma_1x}F^1(x,y)dy\
dx\tilde{f}_{\sigma_1\widetilde{W}^1_T+\sigma_2\widetilde{W}^2_T}(z)dz\\[2ex]
&-e^{-rT}\frac{K}{S^2_0}e^{-(r-\frac{1}{2}\sigma^2_2)T}\int_{\tilde{d}}^{+\infty}\int_{-\infty}^{+\infty}\int_{w(c,z)}^{+\infty}e^{-\sigma_2x}F^2(x,y)dy\
dx\tilde{f}_{\sigma_1\widetilde{W}^1_T+\sigma_2\widetilde{W}^2_T}(z)dz,
\end{align*}
where $F^1$, $F^2$ stand for the density functions of the two
dimensional normal distributions $N_2(M^1(z),\Sigma^1(z))$,
$N_2(M^2(z),\Sigma^2(z))$ respectively.

\subsection{Outperformance option}
The problem is studied for
\begin{gather*}
H=\left(\max\{S^1_T,S^2_T\}-K\right)^{+}, \quad K>0.
\end{gather*}
Let us notice that {\small
\begin{align}\label{wzor 1 OutOp}
\left\{e^{A_1W^1_T+A_2W^2_T+BT}\geq c(S^1_T-K)\right\}&\hskip-0.5ex=\hskip-0.5ex\left\{A_2
W^2_T\geq
v_1(c,W^1_T)\right\}\hskip-0.5ex=\hskip-0.5ex\left\{A_2\widetilde{W}^2_T\geq w_1(c,\widetilde{W}^1_T)\right\}\\[2ex]\label{wzor 2 OutOp}
\left\{e^{A_1W^1_T+A_2W^2_T+BT}\geq c(S^2_T-K)\right\}&\hskip-0.5ex=\hskip-0.5ex\left\{A_1
W^1_T\geq v_2(c,W^2_T)\right\}\hskip-0.5ex=\hskip-0.5ex\left\{A_1\widetilde{W}^1_T\geq
w_2(c,\widetilde{W}^2_T)\right\},
\end{align}
} where{\small
\begin{align*}
v_1(c,x)&:=\ln\left[cS^1_0e^{(\alpha_1-\frac{1}{2}\sigma_1^2)T+\sigma_1
x}\right]-A_1x-BT,\\[2ex]
v_2(c,y)&:=\ln\left[cS^2_0e^{(\alpha_2-\frac{1}{2}\sigma_2^2)T+\sigma_2
y}\right]-A_2y-BT,\\[2ex]
w_1(c,x)&:=\ln\left(ce^{-A_1x+(A_1\frac{\alpha_1-r}{\sigma_1}-B)T}(S^1_0e^{(r-\frac{1}{2}\sigma_1^2)T+\sigma_1x}-K)\right)+\frac{\alpha_2-r}{\sigma_2}T,\\[2ex]
w_2(c,y)&:=\ln\left(ce^{-A_2y+(A_2\frac{\alpha_2-r}{\sigma_2}-B)T}(S^2_0e^{(r-\frac{1}{2}\sigma_2^2)T+\sigma_2y}-K)\right)+\frac{\alpha_1-r}{\sigma_1}T.
\end{align*}}
By \eqref{wzor S^1>K}, \eqref{wzor S^2>K}, \eqref{wzor S^1>S^2} we
have 
\begin{align*}
&\Psi_1(c)=P\left(e^{A_1W^1_T+A_2W^2_T+BT}\geq c(S^1_T\vee
S^2_T-K)^{+}\right)=P\Big(e^{A_1W^1_T+A_2W^2_T+BT}\geq c\\[1ex]
&\cdot(S^1_T-K)\mid W^1_T\geq
a_1,\sigma_1W^1_T-\sigma_2W^2_T\geq b\Big)P(W^1_T\geq
a_1,\sigma_1W^1_T-\sigma_2W^2_T\geq b)\\[1ex]
&+P\left(e^{A_1W^1_T+A_2W^2_T+BT}\geq 0\mid W^1_T<
a_1,\sigma_1W^1_T-\sigma_2W^2_T\geq b\right)P(W^1_T<
a_1,\\[1ex]
&\phantom{+}\sigma_1W^1_T-\sigma_2W^2_T\geq b)
+P(e^{A_1W^1_T+A_2W^2_T+BT}\geq c(S^2_T-K)\mid W^2_T\geq
a_2,\\[1ex]
&\phantom{+}\sigma_1W^1_T-\sigma_2W^2_T< b)P(W^2_T\geq
a_2,\sigma_1W^1_T-\sigma_2W^2_T< b)
+P(e^{A_1W^1_T+A_2W^2_T+BT}\\[1ex]
&\phantom{+}\geq 0\mid W^2_T<a_2,\sigma_1W^1_T-\sigma_2W^2_T< b)P(W^2_T<
a_2,\sigma_1W^1_T-\sigma_2W^2_T< b).
\end{align*}
By \eqref{wzor 1 OutOp}, \eqref{wzor 2 OutOp} we have 
\begin{align*}
&\Psi_1(c)=P\left(A_2W^2_T\geq v_1(c,W^1_T)\mid W^1_T\geq
a_1,\sigma_1W^1_T-\sigma_2W^2_T\geq b\right)\\[1ex]
&\cdot P(W^1_T\geq
a_1,\sigma_1W^1_T-\sigma_2W^2_T\geq b)
+P(W^1_T<
a_1,\sigma_1W^1_T-\sigma_2W^2_T\geq b)\\[1ex]
&+P\left(A_1W^1_T\geq v_2(c,W^2_T)\mid W^2_T\geq
a_2,\sigma_1W^1_T-\sigma_2W^2_T< b\right)\\[1ex]
&\cdot P(W^2_T\geq
a_2,\sigma_1W^1_T-\sigma_2W^2_T< b)+P(W^2_T< a_2,\sigma_1W^1_T-\sigma_2W^2_T< b).
\end{align*}
Let $m_1(y,z), m_2(x,z), \sigma_1(y,z), \sigma_2(x,z)$ be the
means and variances of the conditional distributions
\begin{align*}
\mathcal{L}(A_1W^1_T\mid
W^2_T=y,\sigma_1W^1_T-\sigma_2W^2_T=z)&=N(m_1(y,z),\sigma_1(y,z)),\\[2ex]
\mathcal{L}(A_2W^2_T\mid
W^1_T=x,\sigma_1W^1_T-\sigma_2W^2_T=z)&=N(m_2(x,z),\sigma_2(x,z)),
\end{align*}
given by \eqref{wzor na srednia i wariancje waunkowa ogolny}. Then
we have
\begin{align*}
\Psi_1(c)&=\int_{a_1}^{+\infty}\int_{b}^{+\infty}\Phi\left(\frac{m_2(x,z)-v_1(c,x)}{\sqrt{\sigma_2(x,z)}}\right)f_{W^1_T,\sigma_1W^1_T-\sigma_2W^2_T}(x,z)dzdx\\[2ex]
&+\int_{-\infty}^{a_1}\int_{b}^{+\infty}f_{W^1_T,\sigma_1W^1_T-\sigma_2W^2_T}(x,z)dzdx\\[2ex]
&+\int_{a_2}^{+\infty}\int_{-\infty}^{b}\Phi\left(\frac{m_1(y,z)-v_2(c,y)}{\sqrt{\sigma_1(y,z)}}\right)f_{W^2_T,\sigma_1W^1_T-\sigma_2W^2_T}(y,z)dzdy\\[2ex]
&+\int_{-\infty}^{a_2}\int_{-\infty}^{b}f_{W^2_T,\sigma_1W^1_T-\sigma_2W^2_T}(y,z)dzdy.
\end{align*}
By \eqref{wzor S^1>K}, \eqref{wzor S^2>K}, \eqref{wzor S^1>S^2},
\eqref{wzor 1 OutOp}, \eqref{wzor 2 OutOp} we have 
\begin{align*}
&\Psi_2(c)=e^{-rT}\mathbf{\tilde{E}}\left((S^1_T\vee
S^2_T-K)^{+}\mathbf{1}_{A_c}\right)\\[1ex]
&=e^{-rT}\mathbf{\tilde{E}}\left((S^1_T-K)\mathbf{1}_{A_c}\mid S^1_T\geq K, S^1_T\geq S^2_T\right)\widetilde{P}(S^1_T\geq K, S^1_T\geq S^2_T)\\[1ex]
&+e^{-rT}\mathbf{\tilde{E}}\left((S^2_T-K)\mathbf{1}_{A_c}\mid S^2_T\geq K, S^1_T< S^2_T\right)\widetilde{P}(S^2_T\geq K, S^1_T< S^2_T)\\[1ex]
&=e^{-rT}\mathbf{\tilde{E}}\left((S^1_T-K)\mathbf{1}_{\{A_2\widetilde{W}^2_T\geq w_1(c,\widetilde{W}^1_T)\}}\mid \widetilde{W}^1_T\geq \tilde{a}_1, \sigma_1\widetilde{W}^1_T-\sigma_2\widetilde{W}^2_T\geq \tilde{b}\right)\\[1ex]
&\phantom{+}\cdot\widetilde{P}(\widetilde{W}^1_T\geq \tilde{a}_1, \sigma_1\widetilde{W}^1_T-\sigma_2\widetilde{W}^2_T\geq \tilde{b})\\[1ex]
&+e^{-rT}\mathbf{\tilde{E}}\left((S^2_T-K)\mathbf{1}_{\{A_1\widetilde{W}^1_T\geq
w_2(c,\widetilde{W}^2_T)\}}\mid \widetilde{W}^2_T\geq \tilde{a}_2,
\sigma_1\widetilde{W}^1_T-\sigma_2\widetilde{W}^2_T<
\tilde{b}\right)\\[1ex]
&\phantom{+}\cdot\widetilde{P}(\widetilde{W}^2_T\geq \tilde{a}_2,
\sigma_1\widetilde{W}^1_T-\sigma_2\widetilde{W}^2_T< \tilde{b}).
\end{align*}
Let $m_1(y,z), \sigma_1(y,z)$ and $m_2(x,z), \sigma_2(x,z)$ denote
means and variances of the conditional distributions
\begin{align*}
\tilde{\mathcal{L}}\left(A_1\widetilde{W}^1_T\mid\widetilde{W}^2_T,
\sigma_1\widetilde{W}^1_T-\sigma_2\widetilde{W}^2_T\right)&=N\left(m_1(y,z),
\sigma_1(y,z)\right),\\[2ex]
\tilde{\mathcal{L}}\left(A_2\widetilde{W}^2_T\mid\widetilde{W}^1_T,
\sigma_1\widetilde{W}^1_T-\sigma_2\widetilde{W}^2_T\right)
&=N\left(m_2(x,z), \sigma_2(x,z)\right).
\end{align*}
Finally we obtain
\begin{align*}
\Psi_2(c)&=e^{-rT}\int_{\tilde{a}_1}^{+\infty}\int_{\tilde{b}}^{+\infty}\left(S^1_0e^{(r-\frac{1}{2}\sigma_1^2)T+\sigma_1x}-K\right)\Phi\left(\frac{m_2(x,z)-w_1(c,x)}{\sqrt{\sigma_2(x,z)}}\right)\\[1ex]
&\phantom{+}\cdot\tilde{f}_{\widetilde{W}^1_T,\sigma_1\widetilde{W}^1_T-\sigma_2\widetilde{W}^2_t}(x,z)dz dx\\[1ex]
&+e^{-rT}\int_{\tilde{a}_2}^{+\infty}\int_{-\infty}^{\tilde{b}}\left(S^2_0e^{(r-\frac{1}{2}\sigma_2^2)T+\sigma_1y}-K\right)\Phi\left(\frac{m_1(y,z)-w_2(c,y)}{\sqrt{\sigma_1(y,z)}}\right)\\[1ex]
&\phantom{+}\cdot\tilde{f}_{\widetilde{W}^2_T,\sigma_1\widetilde{W}^1_T-\sigma_2\widetilde{W}^2_t}(y,z)dz
dy.
\end{align*}

\subsection{Spread option}

The payoff is of the form
\begin{gather*}
H=\left(S^1_T-S^2_T-K\right)^{+}, \quad K>0.
\end{gather*}
For any $y\in\mathbb{R}$
\begin{gather}
\{S^1_T-S^2_0e^{(\alpha_2-\frac{1}{2}\sigma_2^2)T+\sigma_2y}-K\geq
0\}=\{W^1_T\geq e(y)\},
\end{gather}
where
\begin{gather*}
e(y):=\frac{1}{\sigma_1}\left(\ln\left[\frac{1}{S^1_0}(S^2_0e^{(\alpha_2-\frac{1}{2}\sigma_2^2)T+\sigma_2y})\right]-\left(\alpha_1-\frac{1}{2}\sigma_1^2\right)T\right).
\end{gather*}
\begin{align*}
&\Psi_1(c)=P\left(e^{A_1W^1_T+A_2W^2_T+BT}\geq
c(S^1_T-S^2_T-K)^{+}\right)\\[1ex]
&=\int_{-\infty}^{+\infty}P\left(e^{A_1W^1_T+A_2W^2_T+BT}\geq
c(S^1_T-S^2_T-K)^{+}\mid W^2_T=y\right)f_{W^2_T}(y)dy\\[1ex]
&=\int_{-\infty}^{+\infty}P\left(e^{A_1W^1_T+A_2W^2_T+BT}\geq
c(S^1_T-S^2_T-K), W^1_T\geq e(y)\mid W^2_T=y\right)\\[1ex]
&\phantom{+}\cdot f_{W^2_T}(y)dy+\int_{-\infty}^{+\infty}P\left(e^{A_1W^1_T+A_2W^2_T+BT}\geq 0,
W^1_T<e(y)\mid W^2_T=y\right)\\[1ex]
&\phantom{+}\cdot f_{W^2_T}(y)dy
=\int_{-\infty}^{+\infty}\hskip-1exP\Big(e^{A_2y+BT}e^{A_1W^1_T}-cS^1_0e^{(\alpha_1-
\frac{1}{2}\sigma_1^2)T}e^{\sigma_1W^1_T}\geq
-c(S^2_T+K), \\[1ex]&\phantom{+}W^1_T\geq e(y)\mid
W^2_T=y\Big)f_{W^2_T}(y)dy
+\int_{-\infty}^{+\infty}\hskip-1exP\left(W^1_T<e(y)\mid
W^2_T=y\right)f_{W^2_T}(y)dy.
\end{align*}
Let
\begin{gather*}
\mathcal{L}(W^1_T\mid W^2_T=y)=N(m(y),\sigma(y)).
\end{gather*}
Then we have
\begin{align*}
\Psi_1(c)&=\int_{-\infty}^{+\infty}\int_{S(c,y)\cap (e(y),
+\infty)}f_{W^1_T\mid W^2_T=y}(x)dx
f_{W^2_T}(y)dy\\[1ex]
&+\int_{-\infty}^{+\infty}\Phi\left(\frac{e(y)-m(y)}{\sqrt{\sigma(y)}}\right)f_{W^2_T}(y)dy,
\end{align*}
where $S(c,y)$ is a set defined $y\in\mathbb{R}$ by
{\small
\begin{gather*}
S(c,y):=\left\{x: e^{A_2y+BT}e^{A_1x}-cS^1_0e^{(\alpha_1-
\frac{1}{2}\sigma_1^2)T}e^{\sigma_1x}\geq
-c(S^2_0e^{(\alpha_2-\frac{1}{2}\sigma_2^2)T+\sigma_2y}+K)\right\}.
\end{gather*}}
For the practical applications it is necessary to find a closed form
of the set $S(c,y)$. In the formulation of the next result we will
use the solutions of the equation
\begin{gather}\label{rownanie w SPREAD}
g(x)= -c(S^2_0e^{(\alpha_2-\frac{1}{2}\sigma_2^2)T+\sigma_2y}+K),
\end{gather}
where $g(x):=e^{A_2y+BT}e^{A_1x}-cS^1_0e^{(\alpha_1-
\frac{1}{2}\sigma_1^2)T}e^{\sigma_1x}$. These solutions can be found
numerically.

\begin{prop}\label{prop SPREAD}
The set $S(c,y)$ is of the form
\begin{enumerate}[a)]
\item if $A_1>\sigma_1$ and
\begin{enumerate}[(i)]
\item if $g(\hat{x})\geq
-c(S^2_0e^{(\alpha_2-\frac{1}{2}\sigma_2^2)T+\sigma_2y}+K)$ then
$S(c,y)=(-\infty,+\infty)$,
\item if $g(\hat{x})\geq
-c(S^2_0e^{(\alpha_2-\frac{1}{2}\sigma_2^2)T+\sigma_2y}+K)$ then
$S(c,y)=(-\infty,x_1)\cup(x_2,+\infty)$, where $x_1<x_2$ are the
unique solutions of \eqref{rownanie w SPREAD}.
\end{enumerate}
Above, $\hat{x}$ stands for
$\frac{1}{\sigma_1-A_1}\ln\left(\frac{A_1e^{A_2y+BT}}{\sigma_1cS^1_0e^{(\alpha_1-
\frac{1}{2}\sigma_1^2)T}}\right)$.

\item if $A_1=\sigma_1$ and
\begin{enumerate}[(i)]
\item $e^{A_2y+BT}\geq cS^1_0e^{(\alpha_1-
\frac{1}{2}\sigma_1^2)T}$ then $S(c,y)=(-\infty, +\infty)$,
\item $e^{A_2y+BT}<cS^1_0e^{(\alpha_1-
\frac{1}{2}\sigma_1^2)T}$ then $S(c,y)=(-\infty, x_0)$, where $x_0$
is a unique solution of \eqref{rownanie w SPREAD},
\end{enumerate}
\item if $A_1<\sigma_1$ then $S(c,y)=(-\infty, x_0)$, where $x_0$ is a
unique solution of \eqref{rownanie w SPREAD}
\end{enumerate}
\end{prop}
{\bf Proof:} $a)$ One can check that $g$ has a minimum at the point
$\hat{x}$ and is decreasing on $(-\infty,\hat{x})$ and increasing on
$(\hat{x},+\infty)$. Hence $(i)$ and $(ii)$ follow.\\
$b)$ The formulas for $S(c,y)$ follows from the simplified form of
the function $g(x)=(e^{A_2y+BT}-cS^1_0e^{(\alpha_1-
\frac{1}{2}\sigma_1^2)T})e^{A_1x}$.\\
$c)$ It can be checked that $g$ is strictly increasing on the set
$\{x: g(x)<0\}$ and that $\lim_{x\rightarrow+\infty}g(x)=-\infty$.
Thus \eqref{rownanie w SPREAD} has a unique solution and the form of
the set $S(c,y)$ follows. \hfill $\square$\\

\noindent Now let us determine $\Psi_2$. One can check that for
$y\in\mathbb{R}$
\begin{gather}
\{S^1_T-S^2_0e^{(r-\frac{1}{2}\sigma_2^2)T+\sigma_2y}-K\geq
0\}=\{\widetilde{W}^1_T\geq f(y)\},
\end{gather}
where
\begin{gather*}
f(y):=\frac{1}{\sigma_1}\left(\ln\left[\frac{1}{S^1_0}(S^2_0e^{(r-\frac{1}{2}\sigma_2^2)T+\sigma_2y})\right]-\left(r-\frac{1}{2}\sigma_1^2\right)T\right).
\end{gather*}
For $y\in\mathbb{R}$ define
\begin{align*}
\tilde{S}(c,y):=\Big\{x:
e^{A_1(x-\frac{\alpha_1-r}{\sigma_1}T)+A_2(y-\frac{\alpha_2-r}{\sigma_2}T)+BT}
\geq
&c\Big(S^1_0e^{(r-\frac{1}{2}\sigma_1^2)T+\sigma_1x}\\[1ex]
&-S^2_0e^{(r-\frac{1}{2}\sigma_2^2)T+\sigma_2y}-K\Big)\Big\}.
\end{align*}
\noindent
Then we have
\begin{align*}
&\Psi_2(c)=e^{-rT}{\mathbf{\tilde{E}}}\left[(S^1_T-S^2_T-K)^{+}\mathbf{1}_{A_c}\right]\\[1ex]
&=e^{-rT}\int_{-\infty}^{+\infty}{\mathbf{\tilde{E}}}\left[(S^1_T-S^2_T-K)^{+}\mathbf{1}_{A_c}\mid \widetilde{W}^2_T=y\right]\tilde{f}_{\widetilde{W}^2_T}(y)dy\\[1ex]
&=e^{-rT}\int_{-\infty}^{+\infty}{\mathbf{\tilde{E}}}\left[(S^1_T-S^2_T-K)\mathbf{1}_{A_c}\mathbf{1}_{\{\widetilde{W}^1_T\geq f(y)\}}\mid \widetilde{W}^2_T=y\right]\tilde{f}_{\widetilde{W}^2_T}(y)dy\\[1ex]
&+e^{-rT}\int_{-\infty}^{+\infty}{\mathbf{\tilde{E}}}\left[(S^1_T-S^2_T-K)^{+}\mathbf{1}_{A_c}\mathbf{1}_{\{\widetilde{W}^1_T< f(y)\}}\mid \widetilde{W}^2_T=y\right]\tilde{f}_{\widetilde{W}^2_T}(y)dy\\[1ex]
&=e^{-rT}\int_{-\infty}^{+\infty}\int_{\tilde{S}(c,y)\cap(f(y),+\infty)}
\left(S^1_0e^{(r-\frac{1}{2}\sigma_1^2)T+\sigma_1x}-S^2_0e^{(r-\frac{1}{2}\sigma_2^2)T+
\sigma_2y}-K\right)\\[1ex]
&\phantom{=}\cdot\tilde{f}_{\widetilde{W}^1_T\mid\widetilde{W}^2_T=y}(x)dx\tilde{f}_{\widetilde{W}^2_T}(y)dy.
\end{align*}
\noindent
The explicit form of the set $\tilde{S}(c,y)$ can be
established in the same way as for $S(c,y)$ in Proposition \ref{prop
SPREAD}.

\end{document}